\documentclass[amsmath,amssymb,reprint]{revtex4-1}
\usepackage{graphicx}
\usepackage{esint}
\usepackage{verbatim}
\usepackage{color}
\usepackage{SIunits}
\usepackage{hyperref}

\newcommand{\Deltalc}{\Delta_{lc}}
\newcommand{\lambdacO}{\lambda_{\text{c}}}

\newcommand{\gzeroO}{g_{\text{0}}}

\newcommand{\omegacO}{\omega_{\text{c}}}
\newcommand{\omegamO}{\omega_{\text{m}}}
\newcommand{\omegamflO}{\omega_{\text{m,fl}}}

\newcommand{\QoO}{Q_{\text{c}}}

\newcommand{\xzpfO}{x_{\text{zpf}}}

\newcommand{\mO}{m_{\text{eff}}}

\newcommand{\vecb}[1]{\mathbf{#1}}

\usepackage{graphicx}
\usepackage{dcolumn}
\usepackage{bm}
\usepackage{epstopdf}
\usepackage{epsfig}
\usepackage{tabls}
\usepackage{color}
\usepackage{verbatim}
\usepackage{amsmath}
\usepackage{color}
\newlength{\myVSpace}
\newsavebox{\astrutbox}
\sbox{\astrutbox}{\rule[-5pt]{0pt}{20pt}}

\begin{document}

\title{Enhanced photon-phonon coupling via dimerization in one-dimensional optomechanical crystals}

\author{Matthew H. Matheny}

\email{matheny@caltech.edu}
\affiliation{Institute for Quantum Information and Matter and Thomas J. Watson, Sr., Laboratory of Applied Physics, California Institute of Technology, Pasadena, CA 91125, USA}

\date{\today}

\begin{abstract}
We show that dimerization of an optomechanical crystal lattice, which leads to folding of the band diagram, can couple flexural mechanical modes to optical fields within the unit cell via radiation pressure. When compared to currently realized crystals, a substantial improvement in the coupling between photons and phonons is found. For experimental verification, we implement a dimerized lattice in a silicon optomechanical nanobeam cavity and measure a vacuum coupling rate of $\gzeroO/2\pi=1.7$~MHz between an optical resonance at $\lambda_{c} = 1545$~nm and a mechanical resonance at $1.14$~GHz.
\end{abstract}

\maketitle

Optomechanical crystals (OMCs)~\cite{eichenfield2009} are periodically structured materials in which optical and acoustic waves are strongly coupled via radiation pressure. For typical solid-state materials, owing to the orders of magnitude difference between the speed of light and sound, near-infrared photons of frequency $\omega/2\pi \sim 200$~THz are matched in wavelength to acoustic waves in the GHz frequency band. Thin-film silicon (Si) OMCs have been used to trap and localize these disparate waves, allowing for a number of proposed experiments in cavity-optomechanics to be realized~\cite{chan2011,safavi2011,safavi2013}.  

An exciting possibility is the creation of an appreciable nonlinearity at the single photon level using patterned dielectric films~\citep{aspelmeyer2014}. However, observing nonlinear photon-phonon interactions requires a vacuum coupling rate $\gzeroO$ larger than the intrinsic optical decay rate $\kappa$~\cite{nunnenkamp2011}.  In addition, the mechanical frequencies must be larger than optical decay rates with  $\omegamO > \kappa/2$, i.e. be 'sideband-resolved'. Currently, sideband-resolved optomechanical systems are two orders of magnitude away from an appreciable nonlinear  interaction~\cite{aspelmeyer2014}, $\gzeroO/\kappa \approx 0.01$. Here, we theoretically show single photon-phonon strong coupling $\gzeroO/\kappa \approx 1$ is possible in optomechanical crystals cavities.

The OMC design which demonstrates the strongest coupling in silicon~\cite{chan2011} has been implemented in materials with a lower index of refraction, such as silicon nitride (Si$_3$N$_4$)~\cite{davanco2014}, aluminum nitride (AlN)~\cite{vainsencher2014}, and Diamond~\cite{burek2016}. In those works the coupling does not exceed $200$~kHz, while Chan, et al.~\cite{chan2012} show a coupling $\gzeroO=1.1$~MHz. This difference arises from the nature of the optomechanical interaction. Chan, et al. find their optomechanical interaction is primarily due to the photoelastic effect, whose matrix element scales as the fourth power of the index of refraction~\citep{chan2012}. This leads to significantly smaller coupling in materials with a lower index. Here we show that the coupling can be significantly improved in a lower index material using the moving boundary interaction, whose matrix element scales as the square of the index.

The optical frequency shift per unit displacement for a moving boundary in a dielectric optical cavity was derived by Johnson, et. al~\cite{Johnson2002}. Combining this with the mechanical zero-point fluctuations, $\xzpfO=\sqrt{\hbar/2\mO\omegamO}$, gives the vacuum coupling rate $g_{0,MB}$, where $\mO$ is the effective mass of the mechanical mode with frequency $\omegamO$. This rate can be written in the form~\citep{eichenfield2009},
\begin{equation}\label{eq:gob}
\begin{aligned}
\begin{split}
g_{0,MB} = &-\sqrt{\frac{\hbar}{8}}*\frac{\omegacO}{\sqrt{\omegamO}}* \\
&\frac {\int_{\partial V} (\vecb{q}(\vecb{r}) \cdot \hat{n})(\Delta\Bar{\Bar{\epsilon}} \vert \vecb{E_{\vert\vert}}\vert^2-\Delta(\overline{\overline{\epsilon^{-1}}}) \vert\vecb{D}_{\bot}\vert^2)d^2r} {\sqrt{\int_V \rho |\vecb{q}(\vecb{r})|^2 d^3r(\vecb{r})\vert)}\int_V \Bar{\Bar{\epsilon}}(\vecb{r})\vert\vecb{E}(\vecb{r})\vert^2d^3r} 
\end{split}
\end{aligned}
\end{equation}
\noindent where $\hat{n}$ is the outward vector normal to the surface of the dielectric boundary, $\bar{\bar{\epsilon}}$ is the dielectric tensor, $\vecb{E}$ ($\vecb{D}$) is the electric (displacement) field, $\partial V$ is the surface of the dielectric structure with volume $V$, and $\vecb{q}(\vecb{r})$ is the unit-normalized mechanical displacement field~\cite{safavi2010}. A similar equation can be expressed for the photoelastic contribution to the optomechanical coupling. Given this equation with fixed material properties, the possible strategies for increasing coupling are: increasing mode overlap, decreasing mode volumes, increasing optical cavity frequency, or decreasing mechanical frequency. In this work, we focus on decreasing mechanical frequencies (which boosts $\xzpfO$) while leaving the other quantities of the equation fixed.

Since the optical intensity profile of the unit cell comprising photonic crystal cavities is typically mirror symmetric about the midpoint, previous OMC designs focused on using fully symmetric extensional-type mechanical modes within the unit cell~\cite{eichenfield2009,chan2012}. However, the lowest frequency eigenmodes usually involve mechanical torsion or flexure. Thus, we outline a method for using flexural modes within the unit cell. 

Engineering coupling between flexural mechanical modes and optical modes in resonant OMC cavities is problematic. Flexural modes are usually not symmetric at the $\Gamma$-point of the band diagram. Thus, they do not couple into symmetric optical modes at the $X$-point. In this work, we design a fully symmetric flexural mode at the $\Gamma$-point via dimerization of the lattice. Thus, we can preserve the mode volumes of the unit cell while decreasing mechanical mode frequencies. In addition, the major contribution to the coupling is from the moving boundary of the dielectric due to mechanical flexure.

Lattice dimerization was first discussed by Peierls~\cite{peierls1955} who predicted an energy gap in the electronic band structure of atomic systems. By doubling the size of the unit cell and breaking the degeneracy (via different hole sizes), we engineer a dimer unit cell with $k$-vector the sum of the two constituent $k$-vectors. If we choose a constituent $k$-vector of alternating $\pm\pi/a$ where $a$ is the lattice constant, we can create a null $k$-vector for the dimer. In essence, we imbue the OMC lattice with a two-'atom' basis of flexing beams. We illustrate how this corresponds to a symmetrized displacement vector in the 1-D lattice.

\begin{figure}[htbp]
\includegraphics[width=8cm]{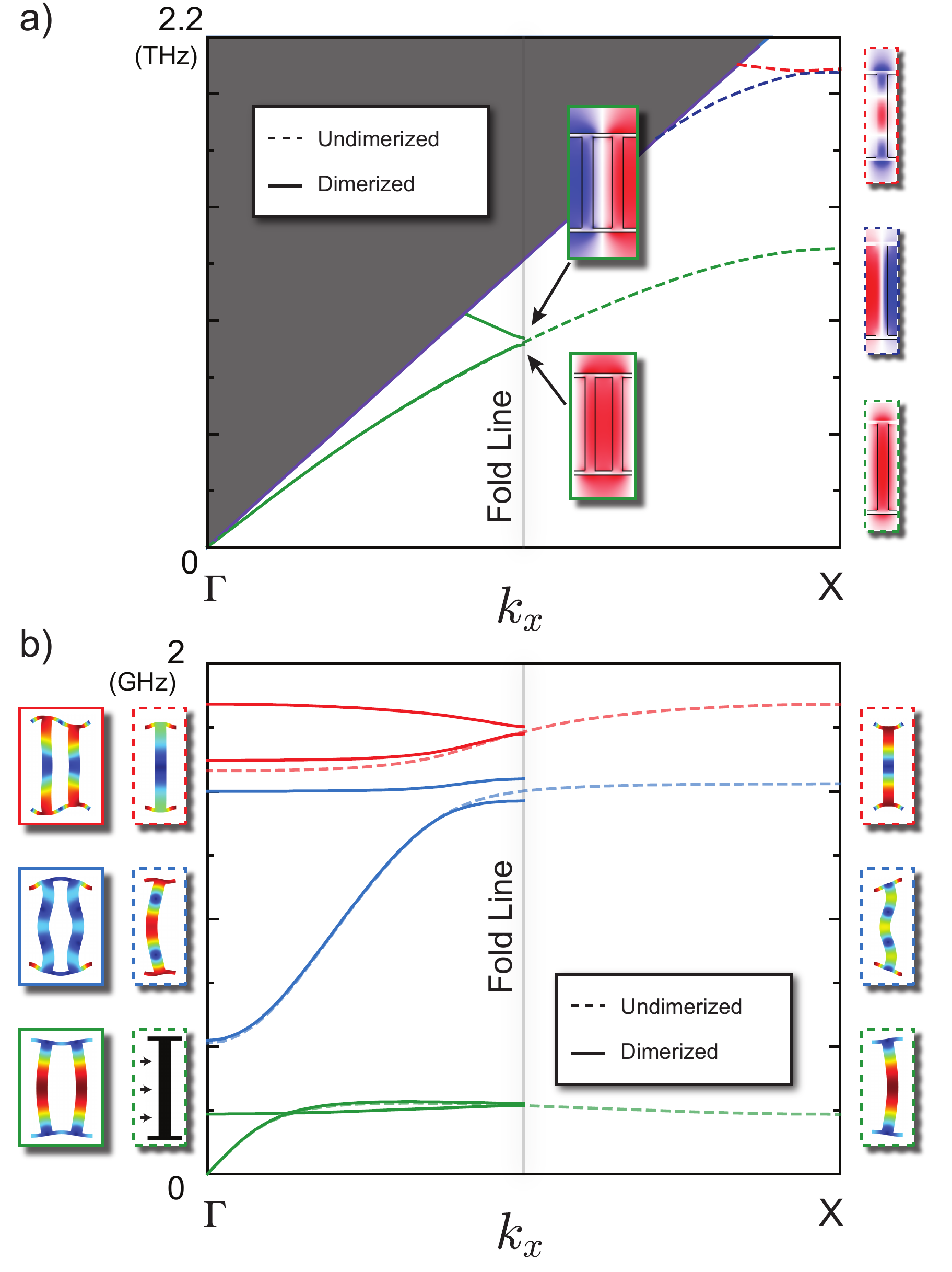}
\caption{For both optical and mechanical band structures the dotted lines are bands for the unit cell before dimerization, the solid lines after dimerization.  The unit cell has dimensions $\lbrace a,w_y,h_x,h_y,t\rbrace = \lbrace 250nm,1500nm,125nm,1300nm,220nm\rbrace$, where $a$ is the lattice constant, $w_y$ is the extent of the body in the y-dimension, and $h_x,h_y$ are the hole dimensions in the x and y directions, respectively. In the dimerized unit cell $\lbrace h_{x1},h_{x2}\rbrace=\lbrace 100nm,150nm\rbrace$. \textbf{a)} Simulated optical band structure of a "ladder" OMC in silicon. Here we show only the modes with electric field (vector) symmetry in the $y$ and $z$ axes. Color in the unit cell plots indicates the value of the electric field in y ($E_{y}(\vecb{r})$). Simulations are performed with the MIT Photonic Bands (MPB) package. \textbf{b)} Mechanical band structure of the same OMC as (a).  We show the acoustic modes with vector displacement symmetry in the $y$ and $z$ axes. Color in the unit cell plots indicates total displacement ($|\vecb{Q}(\vecb{r})|$). Simulations are performed in COMSOL.}
\label{fig:bands}
\end{figure}

We begin the discussion by dimerizing the first OMC design~\cite{eichenfield2009}, which is based on a simple "ladder" structure. Throughout this Letter, the $x$-axis is in the direction of the lattice and the $z$-axis is out of page. In Fig.~\ref{fig:bands}, we show a simulation of the band structure for the "ladder" OMC in silicon, before (dashed lines) and after (solid lines) dimerization. It was found that the largest coupling occured between the $X$-point optical "dielectric" mode (Fig.~\ref{fig:bands}(a) right hand side, green-dashed) and the $\Gamma$-point "breathing" mode of the mechanics (Fig.~\ref{fig:bands}(b) left hand side, red-dashed). These modes exhibit strong overlap; this gives a large photoelastic contribution to the coupling in high index materials. Since the electric field is not designed to be maximum at the boundaries, this type of design does not emphasize optomechanical coupling due to a moving boundary. 

The flexural mechanical modes of the simple "ladder" OMC are the first $X$-point mode (Fig.~\ref{fig:bands}(b), right hand side, green-dashed) and the second $\Gamma$-point mode (Fig.~\ref{fig:bands}(b), left hand side, blue-dashed). These modes do not couple to any of the $X$-point optical modes according to Eqn.~\eqref{eq:gob} due to antisymmetry of the displacement and strain fields in the $x$-axis.
\begin{figure}[htbp]
\includegraphics[width=8cm]{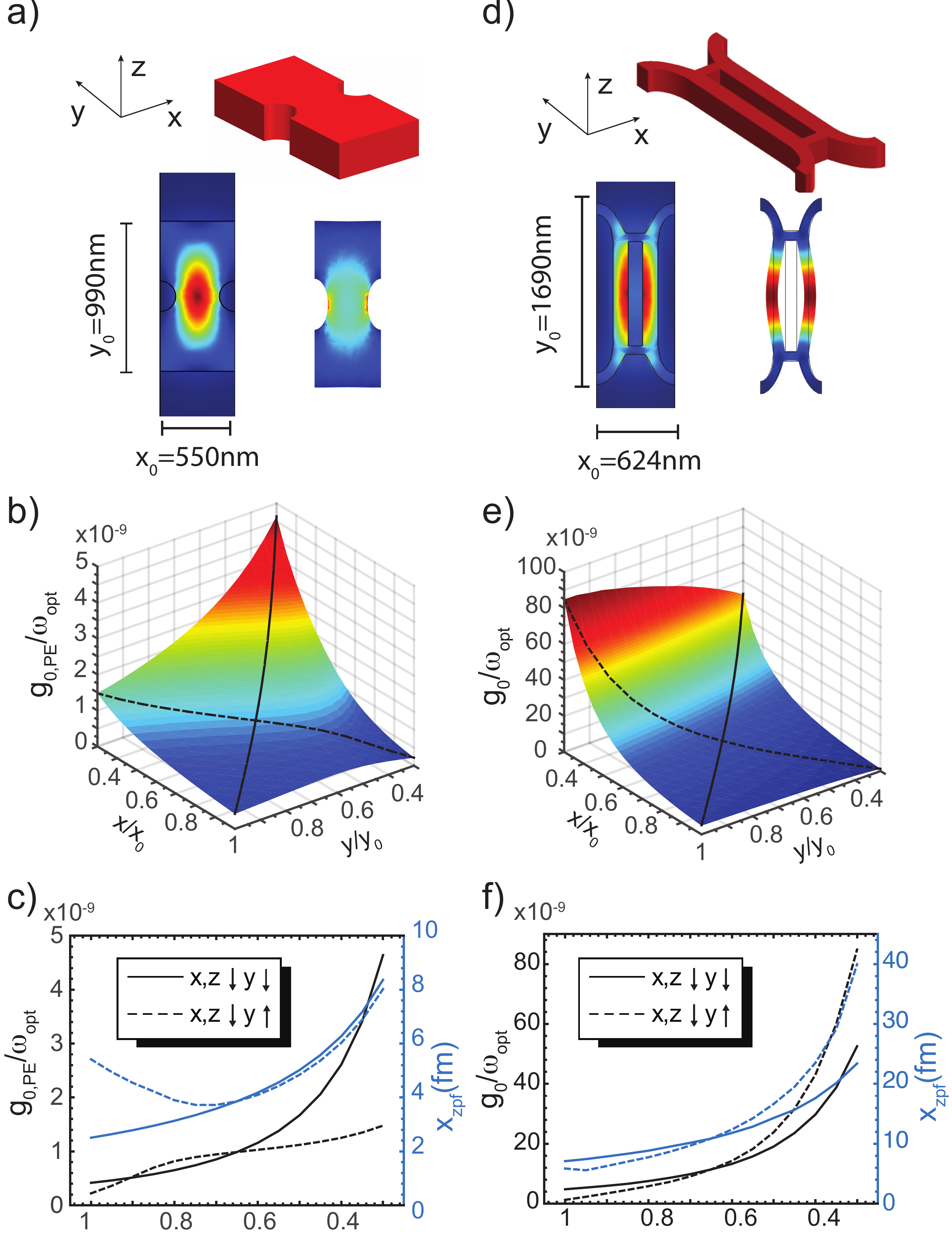}
\caption{Simulated scaling of aluminum nitride unit cell coupling. The coupling is scaled by the optical mode frequency in order to highlight the effect of both the optical/elastic field overlap and enhanced $\xzpfO$ of the new design. In the simulation, since the $z$-axis thickness is chosen by wavelength, and the $x$-axis parameters strongly determine the wavelength, the parameters related to $x$ and $z$ are swept simultaneously. The y-parameters are stepped separately.The initial AlN thickness is set to $330$nm. \textbf{a,d)} Unit cell, optical field intensity (left a,d) of the first dielectric band and mechanical y-axis strain (right a) of the "breathing" mode, and displacement (right d) of the "flexural" mode, respectively. \textbf{b,e)} Optomechanical coupling for the unit cell as different sets of dimensions are scaled. \textbf{c,f)} Contributions along different diagonals of (b,e) to $\gzeroO$. Simulations are performed in COMSOL.}
\label{fig:alnscale}
\end{figure}

However, dimerization can give a symmetric flexural mode (Fig.~\ref{fig:bands}(b) left hand side, green solid line). Also, the lowest "dielectric" mode will now be split into optical modes whose electric field intensity is strongest at different pairs of interior dielectric boundaries. This creates a strong overlap between the electric field intensity and the displacement field.  Essentially, dimerization leads to folding of both band structures, sending $X$-point modes to the $\Gamma$-point of the new lattice, and doubling the number of bands. This is shown in Fig.~\ref{fig:bands}, where the band structure is folded at $k_x = \pi/2a$. The overlap between the new optical modes at the new $X$-point (Fig.~\ref{fig:bands}(a), green solid line) and folded bottom mechanical mode (Fig.~\ref{fig:bands}(b), left hand side, green solid line) now gives a finite coupling. 

Next we study the differences in optomechanical coupling between the "breathing" OMC and the dimerized "flexural" OMC as we scale the unit cell along $x$ and $y$. This analysis emphasizes the benefits of a dimerized design when using lower index materials. Here, we analyze a recently reported OMC~\cite{vainsencher2014} unit cell with the material properties of AlN. In Fig.~\ref{fig:alnscale}(a,b,c) we show simulations of the optomechanical coupling via the photoelastic effect between the 1st optical dielectric mode and the "breathing" mechanical mode, similar to previous designs~\cite{chan2011,vainsencher2014,davanco2014}. In Fig.~\ref{fig:alnscale}(d,e,f) we show simulations for the dimerized lattice, where the unit cell degeneracy has been strongly broken to generate a large optical band gap useful for making high quality cavities. The simulation parameters are initially set so that the optical wavelength of the $\Gamma$-point eigenmode of the OMC unit cell is $1550$nm. The bare optomechanical couplings are scaled by the optical frequency found in simulation, which removes the contribution due to $\omega_c$ from Eqn. \ref{eq:gob}. Thus, the figure highlights the contributions from field overlap and $\xzpfO$.

Fig.~\ref{fig:alnscale}(a) shows the structure of the "breathing" unit cell along with the optical mode's electric field energy and mechanical mode's $y$-strain. Fig.~\ref{fig:alnscale}(b) shows the photoelastic coupling as the $x$ and $z$ parameters are changed in one axis of the surface plot, with the $y$ parameters changed along the other axis of the surface plot. Note that the moving boundary coupling in the "breathing" mode can either add or subtract from the overall coupling in this type of unit cell, and is thus not included. This does not detract from the point of the analysis, which is primarily concerned with the effects of $\xzpfO$. The value found for the photoelastic coupling within the unit cell is consistent with previous work~\cite{vainsencher2014}. In Fig.~\ref{fig:alnscale}(c), we show the individual contributions from $g_0/\omega_{opt}$ and $\xzpfO$ along two different diagonals of the surface plot. 

Fig.~\ref{fig:alnscale}(d) shows the "flexural" OMC structure with electric field energy and total displacement. The overall coupling (photoelastic and moving boundary contributions) is plotted in Fig.~\ref{fig:alnscale}(e), where the photoelastic part only adds to, but is less than 1$\%$ of, the coupling due to the moving boundary. The contributions to this coupling are shown in Fig.~\ref{fig:alnscale}(f) moving along two different diagonals.

Comparing Fig.~\ref{fig:alnscale}(b,e) shows several key differences. First, the dimerized OMC has larger coupling throughout the parameter space. Second, the scaling of the coupling is different. Moving along the solid line shows the simulated coupling when $x$, $y$, and $z$ are all scaled down. In the "breathing" mode, there is an increase in both $\gzeroO/\omega_{opt}$ and $\xzpfO$ as a function of 1/$\sqrt[3]{V}$, as expected. This is also true in the plot for the "flexural" OMC, but there is a larger increase moving along the other diagonal, i.e. along the dashed line. Along that line, the $y$ parameters are scaled up, while the $x$ and $z$ axis parameters are scaled down. The differences between the coupling in the  two types of OMCs is due to the difference in the scaling of $\xzpfO$. Decreasing the $x$ axis parameters by a constant $\zeta$ and increasing the $y$ axis parameters by the same constant leaves the mass fixed. However, the "flexural" mode will decrease in frequency by a factor  $\propto \zeta^3$, while the "breathing" mode will decrease by a factor $\propto \zeta$. This leads to larger $\xzpfO$ for the "flexural" modes. 

The simulation for the "flexural" mode can be compared against the "breathing" mode for optical wavelengths near $1550$nm ($x_0,y_0 = 1$). The "breathing" mode simulation yields $\omega_{m,br}/2\pi=4.45 GHz$, $g_{0,PE}/2\pi=135kHz$ while the flexural mode gives $\omegamflO/2\pi=1.39GHz$, $\gzeroO/2\pi=928kHz$. This is nearly a $7\times$ improvement in the coupling. This difference becomes even greater if the lattice is scaled. 

\begin{table}[h]
\caption{\label{tab:1}Four simulated cavities using the dimerized lattice from Fig.~\ref{fig:alnscale}(d). Simulations are performed in COMSOL.}
\begin{ruledtabular}
\begin{tabular} {lllll}
Material                                  & Si   & Si   & GaAs & SiC  \\ 
Index                                     & 3.48 & 3.48 & 3.42 & 2.59 \\ 
Lattice constant, $a$ (nm)                     & 480  & 380  & 275  & 190  \\ 
Small hole, $w$ (nm)                           & 70   & 60   & 40   & 15   \\ 
$\lambdacO$ (nm)                           & 1570 & 1280 & 950  & 545  \\ 
$\omegamO/2\pi$ (GHz)                & 1.07 & 0.65 & 0.73 & 0.65 \\ 
$\xzpfO$ (fm)                               & 7.1  & 10   & 11   & 16   \\ 
$\gzeroO/2\pi$ (MHz)                           & 1.7  & 4.0  & 8.8  & 26   \\ 
$\gzeroO/\kappa$                                & 0.081 & 0.17  & 0.27  & 0.47 \\

\end{tabular}
\end{ruledtabular}
\end{table}

Cavities using this type of OMC can achieve single photon-phonon strong coupling. We report four simulations in Table \ref{tab:1} for an optomechanical cavity constructed from a dimerized crystal similar to Fig. \ref{fig:alnscale}(d). In the first two columns we use the material properties of silicon, with optical modes near $1550$nm and $1300$nm made from the second optical band of Fig.~\ref{fig:bands}. In the third and fourth column, we use the material properties of gallium arsenide (GaAs) and silicon carbide (SiC) with a design frequency near their respective absorption band edges. 

In the first half of the table we show parameters for the cavity design and fabrication, demonstrating the possibility of such devices using current technologies. Next, we give the designed optical wavelength $\lambdacO$, mechanical frequency $\omegamO$ and zero-point fluctuations $\xzpfO$. Finally, we give the coupling rate $\gzeroO/2\pi$, in addition to the strong coupling parameter $\gzeroO/\kappa$. The quality factors of the optical cavity were assumed to be limited to $\QoO=10^7$ (a high, but realizable $\QoO$ in silicon photonic crystals~\cite{sekoguchi2014}). All of our simulations show a radiation-limited quality factor greater than this value. Thus, the last column shows that strong coupling is possible in this type of OMC given a $\QoO>10^7$~\cite{sekoguchi2014}.

\begin{figure}[htbp]
\includegraphics[width=8cm]{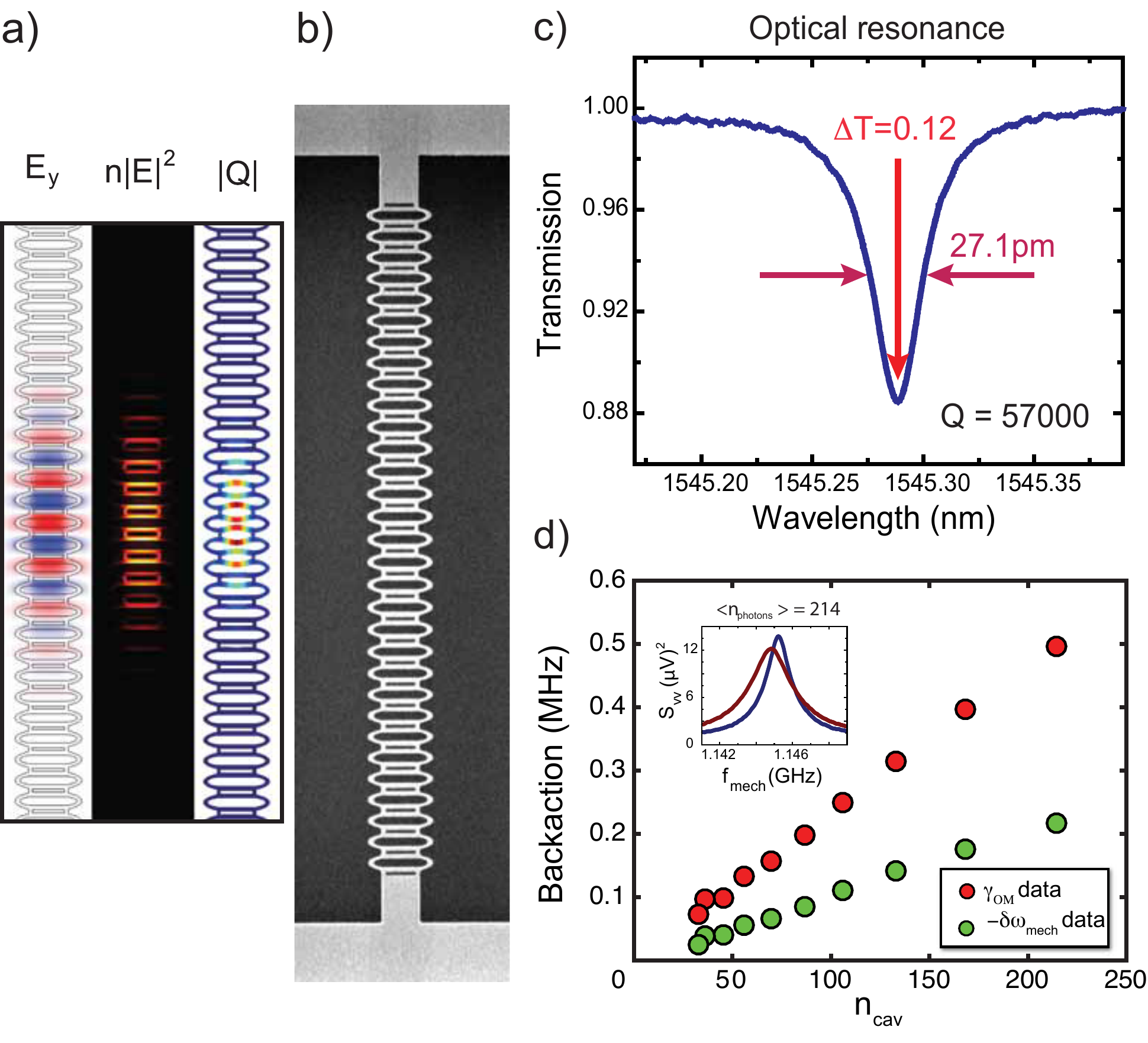}
\caption{\textbf{a)} Simluated mode profiles for the OMC cavity design from the first column of Table~\ref{tab:1}.  From left to right: electric field $y$-polarization ($E_{y}(\vecb{r})$), electric energy density ($n|\vecb{E}(\vecb{r})|^2$), and displacement field ($|\vecb{Q}(\vecb{r})|$). \textbf{b)} SEM image of a fabricated Si OMC cavity with the dimerized lattice. \textbf{c)} Optical transmission spectrum of the OMC cavity. \textbf{d)} Change in mechanical frequency and linewidth as a function of photon number for laser detuning $\Deltalc \approx -0.4\kappa$.  Inset: optically transduced mechanical spectra (after amplifier and photodetector) with detuning set to $\Deltalc \approx -0.4\kappa$ (red curve) and $\Deltalc \approx.4\kappa$ (blue curve).}
\label{fig:exp}
\end{figure}

Finally, we experimentally demonstrate the OMC cavity simulated by the first column of Table \ref{tab:1}. The cavity has $32$ overall unit cells ($12$ for the defect and $20$ for the mirrors). In Fig.~\ref{fig:exp}(a), we show the simulation for the $y$-electric field, the total electric field energy, and the mechanical displacement. The fabricated structure is shown in the scanning electron micrograph in Fig.~\ref{fig:exp}(b). This device is probed with a tapered fiber in transmission mode in a nitrogen environment at room temperature~\citep{eichenfield2009}. The experimental result of our device is shown in Fig.~\ref{fig:exp}(c,d). Note that $\QoO$ (from Fig.~\ref{fig:exp}(c)) is lower than found in other silicon photonic crystal resonators, and is most likely due to inexact matching between design parameters and device parameters. The fabrication can be iterated and improved to give much higher quality factors~\cite{chan2012}.

We probe the mechanical resonances by setting our laser detuning $\Deltalc \approx \pm.4\kappa$ from the optical frequency  (Fig.~\ref{fig:exp}(c)), and measuring the modulations of the transmitted signal~\cite{eichenfield2009}. With laser detuning set at these points on the "red" and "blue" side of the spectrum, we see data as in the inset of Fig.~\ref{fig:exp}(d), which  is in agreement with the simulation for $\omegamO$ up to $\approx 6\%$.

The perturbation to the mechanical frequency and linewidth from the laser pump is due to the imaginary and real parts of the optomechanical backaction, respectively. As can be seen in the main plot, the data at lower input power was subject to more drift in the laser detuning. For every input power we extract the exact laser detuning from the ratio of the imaginary and real parts of the backaction.  Then, for each point, we surmise the occupation of the optical cavity using the incident power and the laser detuning. Thus, the plot represents the extracted occupation factor against real and imaginary parts to the backaction with the red and green circles, respectively. The coupling is then extracted from these data~\citep{safavi2013noise}. This gives $g_{0,exp,real}=1.75 \pm .05 MHz$ and $g_{0,exp,imag}=1.74 \pm .05 MHz$, which agrees with the simulation from Table \ref{tab:1}.

We have shown that dimerized flexural OMC cavities have larger rates of coupling than previously achieved, especially for materials with a low index of refraction.  By using large bandgap materials, very large couplings are achievable with high quality factors, as shown in the analysis of Fig.~\ref{fig:alnscale} and Table~\ref{tab:1}. Also, single photon strong coupling can be engineered in OMC cavities with currently demonstrated photonic crystal quality factors. The principles used in this work can be extended to other types of mechanical modes with odd symmetry, such as torsional or shear modes. They can also be used in two dimensional crystals where mechanical mode symmetry is odd in both dimensions. 

\begin{acknowledgments} 
We thank Oskar Painter for critical advice, support, and use of laboratory and cleanroom facilities. We also thank Sean Meehan, Joseph Redford, Justin Cohen and Mahmood Kalaee for helpful discussions.  This work was supported by the ONR MURI QOMAND program (award no. N00014-15-1-2761), the Institute for Quantum Information and Matter, an NSF Physics Frontiers Center with support of the Gordon and Betty Moore Foundation, and the Kavli Nanoscience Institute at Caltech.
\end{acknowledgments}

\bibliographystyle{apsrev}
\bibliography{Bibliography}

\begin{thebibliography}{15}
\expandafter\ifx\csname natexlab\endcsname\relax\def\natexlab#1{#1}\fi
\expandafter\ifx\csname bibnamefont\endcsname\relax
  \def\bibnamefont#1{#1}\fi
\expandafter\ifx\csname bibfnamefont\endcsname\relax
  \def\bibfnamefont#1{#1}\fi
\expandafter\ifx\csname citenamefont\endcsname\relax
  \def\citenamefont#1{#1}\fi
\expandafter\ifx\csname url\endcsname\relax
  \def\url#1{\texttt{#1}}\fi
\expandafter\ifx\csname urlprefix\endcsname\relax\def\urlprefix{URL }\fi
\providecommand{\bibinfo}[2]{#2}
\providecommand{\eprint}[2][]{\url{#2}}

\bibitem[{\citenamefont{Eichenfield et~al.}(2009)\citenamefont{Eichenfield,
  Chan, Camacho, Vahala, and Painter}}]{eichenfield2009}
\bibinfo{author}{\bibfnamefont{M.}~\bibnamefont{Eichenfield}},
  \bibinfo{author}{\bibfnamefont{J.}~\bibnamefont{Chan}},
  \bibinfo{author}{\bibfnamefont{R.~M.} \bibnamefont{Camacho}},
  \bibinfo{author}{\bibfnamefont{K.~J.} \bibnamefont{Vahala}},
  \bibnamefont{and} \bibinfo{author}{\bibfnamefont{O.}~\bibnamefont{Painter}},
  \bibinfo{journal}{Nature} \textbf{\bibinfo{volume}{462}}, \bibinfo{pages}{78}
  (\bibinfo{year}{2009}).

\bibitem[{\citenamefont{Chan et~al.}(2011)\citenamefont{Chan, Alegre,
  Safavi-Naeini, Hill, Krause, Gr{\"o}blacher, Aspelmeyer, and
  Painter}}]{chan2011}
\bibinfo{author}{\bibfnamefont{J.}~\bibnamefont{Chan}},
  \bibinfo{author}{\bibfnamefont{T.~M.} \bibnamefont{Alegre}},
  \bibinfo{author}{\bibfnamefont{A.~H.} \bibnamefont{Safavi-Naeini}},
  \bibinfo{author}{\bibfnamefont{J.~T.} \bibnamefont{Hill}},
  \bibinfo{author}{\bibfnamefont{A.}~\bibnamefont{Krause}},
  \bibinfo{author}{\bibfnamefont{S.}~\bibnamefont{Gr{\"o}blacher}},
  \bibinfo{author}{\bibfnamefont{M.}~\bibnamefont{Aspelmeyer}},
  \bibnamefont{and} \bibinfo{author}{\bibfnamefont{O.}~\bibnamefont{Painter}},
  \bibinfo{journal}{Nature} \textbf{\bibinfo{volume}{478}}, \bibinfo{pages}{89}
  (\bibinfo{year}{2011}).

\bibitem[{\citenamefont{Safavi-Naeini et~al.}(2011)\citenamefont{Safavi-Naeini,
  Alegre, Chan, Eichenfield, Winger, Lin, Hill, Chang, and
  Painter}}]{safavi2011}
\bibinfo{author}{\bibfnamefont{A.~H.} \bibnamefont{Safavi-Naeini}},
  \bibinfo{author}{\bibfnamefont{T.~M.} \bibnamefont{Alegre}},
  \bibinfo{author}{\bibfnamefont{J.}~\bibnamefont{Chan}},
  \bibinfo{author}{\bibfnamefont{M.}~\bibnamefont{Eichenfield}},
  \bibinfo{author}{\bibfnamefont{M.}~\bibnamefont{Winger}},
  \bibinfo{author}{\bibfnamefont{Q.}~\bibnamefont{Lin}},
  \bibinfo{author}{\bibfnamefont{J.~T.} \bibnamefont{Hill}},
  \bibinfo{author}{\bibfnamefont{D.~E.} \bibnamefont{Chang}}, \bibnamefont{and}
  \bibinfo{author}{\bibfnamefont{O.}~\bibnamefont{Painter}},
  \bibinfo{journal}{Nature} \textbf{\bibinfo{volume}{472}}, \bibinfo{pages}{69}
  (\bibinfo{year}{2011}).

\bibitem[{\citenamefont{Safavi-Naeini
  et~al.}(2013{\natexlab{a}})\citenamefont{Safavi-Naeini, Gr{\"o}blacher, Hill,
  Chan, Aspelmeyer, and Painter}}]{safavi2013}
\bibinfo{author}{\bibfnamefont{A.~H.} \bibnamefont{Safavi-Naeini}},
  \bibinfo{author}{\bibfnamefont{S.}~\bibnamefont{Gr{\"o}blacher}},
  \bibinfo{author}{\bibfnamefont{J.~T.} \bibnamefont{Hill}},
  \bibinfo{author}{\bibfnamefont{J.}~\bibnamefont{Chan}},
  \bibinfo{author}{\bibfnamefont{M.}~\bibnamefont{Aspelmeyer}},
  \bibnamefont{and} \bibinfo{author}{\bibfnamefont{O.}~\bibnamefont{Painter}},
  \bibinfo{journal}{Nature} \textbf{\bibinfo{volume}{500}},
  \bibinfo{pages}{185} (\bibinfo{year}{2013}{\natexlab{a}}).

\bibitem[{\citenamefont{Aspelmeyer et~al.}(2014)\citenamefont{Aspelmeyer,
  Kippenberg, and Marquardt}}]{aspelmeyer2014}
\bibinfo{author}{\bibfnamefont{M.}~\bibnamefont{Aspelmeyer}},
  \bibinfo{author}{\bibfnamefont{T.~J.} \bibnamefont{Kippenberg}},
  \bibnamefont{and}
  \bibinfo{author}{\bibfnamefont{F.}~\bibnamefont{Marquardt}},
  \bibinfo{journal}{Reviews of Modern Physics} \textbf{\bibinfo{volume}{86}},
  \bibinfo{pages}{1391} (\bibinfo{year}{2014}).

\bibitem[{\citenamefont{Nunnenkamp et~al.}(2011)\citenamefont{Nunnenkamp,
  B{\o}rkje, and Girvin}}]{nunnenkamp2011}
\bibinfo{author}{\bibfnamefont{A.}~\bibnamefont{Nunnenkamp}},
  \bibinfo{author}{\bibfnamefont{K.}~\bibnamefont{B{\o}rkje}},
  \bibnamefont{and} \bibinfo{author}{\bibfnamefont{S.}~\bibnamefont{Girvin}},
  \bibinfo{journal}{Physical review letters} \textbf{\bibinfo{volume}{107}},
  \bibinfo{pages}{063602} (\bibinfo{year}{2011}).

\bibitem[{\citenamefont{Davanco et~al.}(2014)\citenamefont{Davanco, Ates, Liu,
  and Srinivasan}}]{davanco2014}
\bibinfo{author}{\bibfnamefont{M.}~\bibnamefont{Davanco}},
  \bibinfo{author}{\bibfnamefont{S.}~\bibnamefont{Ates}},
  \bibinfo{author}{\bibfnamefont{Y.}~\bibnamefont{Liu}}, \bibnamefont{and}
  \bibinfo{author}{\bibfnamefont{K.}~\bibnamefont{Srinivasan}},
  \bibinfo{journal}{Applied Physics Letters} \textbf{\bibinfo{volume}{104}},
  \bibinfo{pages}{041101} (\bibinfo{year}{2014}).

\bibitem[{\citenamefont{Vainsencher et~al.}(2014)\citenamefont{Vainsencher,
  Bochmann, Peairs, Satzinger, Awschalom, and Cleland}}]{vainsencher2014}
\bibinfo{author}{\bibfnamefont{A.}~\bibnamefont{Vainsencher}},
  \bibinfo{author}{\bibfnamefont{J.}~\bibnamefont{Bochmann}},
  \bibinfo{author}{\bibfnamefont{G.}~\bibnamefont{Peairs}},
  \bibinfo{author}{\bibfnamefont{K.}~\bibnamefont{Satzinger}},
  \bibinfo{author}{\bibfnamefont{D.}~\bibnamefont{Awschalom}},
  \bibnamefont{and} \bibinfo{author}{\bibfnamefont{A.}~\bibnamefont{Cleland}},
  \bibinfo{journal}{Bulletin of the American Physical Society}
  (\bibinfo{year}{2014}).

\bibitem[{\citenamefont{Burek et~al.}(2016)\citenamefont{Burek, Cohen,
  Meenehan, El-Sawah, Chia, Ruelle, Meesala, Rochman, Atikian, Markham
  et~al.}}]{burek2016}
\bibinfo{author}{\bibfnamefont{M.~J.} \bibnamefont{Burek}},
  \bibinfo{author}{\bibfnamefont{J.~D.} \bibnamefont{Cohen}},
  \bibinfo{author}{\bibfnamefont{S.~M.} \bibnamefont{Meenehan}},
  \bibinfo{author}{\bibfnamefont{N.}~\bibnamefont{El-Sawah}},
  \bibinfo{author}{\bibfnamefont{C.}~\bibnamefont{Chia}},
  \bibinfo{author}{\bibfnamefont{T.}~\bibnamefont{Ruelle}},
  \bibinfo{author}{\bibfnamefont{S.}~\bibnamefont{Meesala}},
  \bibinfo{author}{\bibfnamefont{J.}~\bibnamefont{Rochman}},
  \bibinfo{author}{\bibfnamefont{H.~A.} \bibnamefont{Atikian}},
  \bibinfo{author}{\bibfnamefont{M.}~\bibnamefont{Markham}},
  \bibnamefont{et~al.}, \bibinfo{journal}{Optica} \textbf{\bibinfo{volume}{3}},
  \bibinfo{pages}{1404} (\bibinfo{year}{2016}).

\bibitem[{\citenamefont{Chan et~al.}(2012)\citenamefont{Chan, Safavi-Naeini,
  Hill, Meenehan, and Painter}}]{chan2012}
\bibinfo{author}{\bibfnamefont{J.}~\bibnamefont{Chan}},
  \bibinfo{author}{\bibfnamefont{A.~H.} \bibnamefont{Safavi-Naeini}},
  \bibinfo{author}{\bibfnamefont{J.~T.} \bibnamefont{Hill}},
  \bibinfo{author}{\bibfnamefont{S.}~\bibnamefont{Meenehan}}, \bibnamefont{and}
  \bibinfo{author}{\bibfnamefont{O.}~\bibnamefont{Painter}},
  \bibinfo{journal}{Applied Physics Letters} \textbf{\bibinfo{volume}{101}},
  \bibinfo{pages}{081115} (\bibinfo{year}{2012}).

\bibitem[{\citenamefont{Johnson et~al.}(2002)\citenamefont{Johnson, Ibanescu,
  Skorobogatiy, Weisberg, Joannopoulos, and Fink}}]{Johnson2002}
\bibinfo{author}{\bibfnamefont{S.~G.} \bibnamefont{Johnson}},
  \bibinfo{author}{\bibfnamefont{M.}~\bibnamefont{Ibanescu}},
  \bibinfo{author}{\bibfnamefont{M.}~\bibnamefont{Skorobogatiy}},
  \bibinfo{author}{\bibfnamefont{O.}~\bibnamefont{Weisberg}},
  \bibinfo{author}{\bibfnamefont{J.}~\bibnamefont{Joannopoulos}},
  \bibnamefont{and} \bibinfo{author}{\bibfnamefont{Y.}~\bibnamefont{Fink}},
  \bibinfo{journal}{Physical review E} \textbf{\bibinfo{volume}{65}},
  \bibinfo{pages}{066611} (\bibinfo{year}{2002}).

\bibitem[{\citenamefont{Safavi-Naeini and Painter}(2010)}]{safavi2010}
\bibinfo{author}{\bibfnamefont{A.~H.} \bibnamefont{Safavi-Naeini}}
  \bibnamefont{and} \bibinfo{author}{\bibfnamefont{O.}~\bibnamefont{Painter}},
  \bibinfo{journal}{Optics express} \textbf{\bibinfo{volume}{18}},
  \bibinfo{pages}{14926} (\bibinfo{year}{2010}).

\bibitem[{\citenamefont{Peierls}(1955)}]{peierls1955}
\bibinfo{author}{\bibfnamefont{R.~E.} \bibnamefont{Peierls}},
  \emph{\bibinfo{title}{Quantum theory of solids}}, \bibinfo{number}{23}
  (\bibinfo{publisher}{Oxford University Press}, \bibinfo{year}{1955}).

\bibitem[{\citenamefont{Sekoguchi et~al.}(2014)\citenamefont{Sekoguchi,
  Takahashi, Asano, and Noda}}]{sekoguchi2014}
\bibinfo{author}{\bibfnamefont{H.}~\bibnamefont{Sekoguchi}},
  \bibinfo{author}{\bibfnamefont{Y.}~\bibnamefont{Takahashi}},
  \bibinfo{author}{\bibfnamefont{T.}~\bibnamefont{Asano}}, \bibnamefont{and}
  \bibinfo{author}{\bibfnamefont{S.}~\bibnamefont{Noda}},
  \bibinfo{journal}{Optics Express} \textbf{\bibinfo{volume}{22}},
  \bibinfo{pages}{916} (\bibinfo{year}{2014}).

\bibitem[{\citenamefont{Safavi-Naeini
  et~al.}(2013{\natexlab{b}})\citenamefont{Safavi-Naeini, Chan, Hill,
  Gr{\"o}blacher, Miao, Chen, Aspelmeyer, and Painter}}]{safavi2013noise}
\bibinfo{author}{\bibfnamefont{A.~H.} \bibnamefont{Safavi-Naeini}},
  \bibinfo{author}{\bibfnamefont{J.}~\bibnamefont{Chan}},
  \bibinfo{author}{\bibfnamefont{J.~T.} \bibnamefont{Hill}},
  \bibinfo{author}{\bibfnamefont{S.}~\bibnamefont{Gr{\"o}blacher}},
  \bibinfo{author}{\bibfnamefont{H.}~\bibnamefont{Miao}},
  \bibinfo{author}{\bibfnamefont{Y.}~\bibnamefont{Chen}},
  \bibinfo{author}{\bibfnamefont{M.}~\bibnamefont{Aspelmeyer}},
  \bibnamefont{and} \bibinfo{author}{\bibfnamefont{O.}~\bibnamefont{Painter}},
  \bibinfo{journal}{New Journal of Physics} \textbf{\bibinfo{volume}{15}},
  \bibinfo{pages}{035007} (\bibinfo{year}{2013}{\natexlab{b}}).

\end{thebibliography}


\end{document}